# Robust quantum point contact operation of narrow graphene constrictions patterned by AFM cleavage lithography

Péter Kun[1*], Bálint Fülöp[2,3], Gergely Dobrik[1], Péter Nemes-Incze[1], István Endre Lukács[1], Szabolcs Csonka[2,3], Chanyong Hwang[4], Levente Tapasztó[1*]

[1] Institute of Technical Physics and Material Science, Centre for Energy Research, 1121 Budapest, Hungary

[2] Department of Physics, Budapest University of Technology and Economics

[3] MTA-BME "Momentum" Nanoelectronics Research Group, H-1111 Budapest, Hungary

4 Korea Research Institute for Standards and Science, Daejeon 305340, South Korea

Detecting conductance quantization in graphene nanostructures turned out more challenging than expected. The observation of well-defined conductance plateaus through graphene nanoconstrictions so far has only been accessible in the highest quality suspended or h-BN encapsulated devices. However, reaching low conductance quanta in zero magnetic field, is a delicate task even with such ultra-high mobility devices. Here, we demonstrate a simple AFM-based nanopatterning technique for defining graphene constrictions with high precision (down to 10 nm width) and reduced edge-roughness (+/- 1 nm). The patterning process is based on the in-plane mechanical cleavage of graphene by the AFM tip, along its high symmetry crystallographic directions. As-defined, narrow graphene constrictions with improved edge quality enable an unprecedentedly robust QPC operation, allowing the observation of conductance quantization even on standard $SiO_2$/Si substrates, down to low conductance quanta. Conductance plateaus, were observed at $n \times e^2/h$, evenly spaced by $2 \times e^2/h$ (corresponding to $n$ = 3, 5, 7, 9, 11) in the absence of an external magnetic field, while spaced by $e^2/h$ ($n$ = 1, 2, 3, 4, 5, 6) in 8T magnetic field.

* Email: kun@mfa.kfki.hu, tapaszto@mfa.kfki.hu

Due to its unique mechanical, optical, and electronic properties, graphene has been at the forefront of low dimensional materials research for fifteen years, starting with the early discovery of the half-integer quantum Hall effect and non-zero Berry phase[1,2]. Due to its outstanding electronic quality, manifesting itself in particularly high charge carrier mobility[3] and long coherence lengths[4], enormous efforts have been invested into realizing graphene-based quantum devices[5,6,7]. However, one of the most fundamental quantum size effects, the precise detection of charge quantization in graphene nanoconstrictions turned out to be particularly challenging, hindering the implementation of graphene-based quantum point contacts.

Before the discovery of 2D crystals, the research on quantum transport phenomena focused on two-dimensional electron systems in semiconductor nanostructures[8]. Among others, quantum point contacts (QPC) – a narrow constriction between two extended electrically conducting areas – have been realized, conventionally through charge carrier depletion with the help of split-gate electrodes[9,10]. Similarly, in bilayer graphene, a transverse electric field can open a band gap[11], which allows for depletion of the gated area, but the full pinch-off state is hard to achieve due to leakage currents. Recently, by improved device fabrication, this obstacle has been removed, allowing the observation of conductance quantization, unexpected mode crossings[12,13], as well as the interplay between spin-orbit coupling and electron-electron interaction at zero magnetic field[14]. However, the presence of AB-BA stacking domain walls cannot be fully avoided in bilayer samples[15]. Acting as current leakage pathways, they limit the reliability of the bilayer QPC device fabrication process.

By contrast, in the case of monolayer graphene, employing a similar split-gate geometry cannot induce a complete electrical depletion, instead it leads to the formation of p-n junctions, short-circuiting the constriction[16]. Nonetheless, the physical removal of material (etching, cutting) provides a viable solution to overcome this problem, and to realize confinement. In this case, the main factors defining the electronic transport characteristics are the charge carrier mobility in graphene and the structural quality of the constriction edges. A major bottleneck is the extended charge localization area, implying that the detrimental effects of edge disorder are felt even relatively far from the constriction[17]. Even so, signatures of quantized conductance have been reported in graphene constrictions. Tombros *et al*. applied a high current annealing method on suspended graphene[18]. However, similar techniques offer a limited control and information on the width and geometry of the constriction. E-beam lithography combined with etching is the most wide-spread nanofabrication technique for graphene quantum QPCs[19]. Although, e-beam lithography is capable of defining constrictions down to a few tens of nanometers width, the size of the investigated constrictions is typically in the hundred nanometers range. This is required to mitigate the effect of edge disorder[20] that precludes obtaining well-defined QPC characteristics in narrower constrictions. However, to reach lower conductance quanta, in such wide constrictions, the Fermi wavelength of graphene has to be of comparable scale to the constriction width. This imposes highly demanding requirements on the graphene quality for enabling QPC operation. Such requirements can only be met by suspending graphene or encapsulating it between hexagonal boron nitride layers. Only in the best ultra-high mobility devices can the signatures of size quantization be observed, in the form of more or less evenly spaced modulations (kinks) superimposed on the linear conductance[21,22,23,24]. However, in the absence of magnetic field, these plateaus only become well-defined further away from the Dirac point, corresponding to high conductance quanta (typically, $\sigma > 10$ $e^2/h$). Reducing the constriction width is important as it can ensure a more robust QPC operation, allowing the observation of plateaus at lower conductance quanta, and enabling QPC characteristics to persist up to higher temperatures, by increasing the energy separation between transversal modes. However, reducing the constriction width cannot be efficient without improving the edge quality. These are two of the next important steps towards improving the feasibility and reliability of graphene QPCs.

Besides e-beam lithography, scanning probe microscopy-based techniques can also be employed for patterning graphene. For instance, AFM-based local anodic oxidation lithography has been applied on graphene to form nanoribbons, quantum dots or nanorings, by etching, down to 30-nm-wide, insulating trenches [25,26,27]. The precise direct mechanical cutting of graphene by AFM turned out quite difficult, as graphene tends to tear and fold along various directions during the scratching [28]. An underlying thin layer of polymethyl methacrylate (PMMA)[29] has been employed, to provide stronger adhesion to the substrate, and improve the cutting precision. On conducting substrates scanning tunneling microscope lithography can be employed, for defining graphene nanoribbons with nanometre precision [30,31]. However, the conductive substrates required by STM do not allow the direct integration of the patterned graphene structures into electronic devices.

Here, we have developed an improved AFM lithographic method enabling a much more precise direct mechanical patterning (in-plane cleavage) of graphene sheets. This technique enabled us to define graphene nanoconstrictions down to 10 nm width, with edges of improved structural quality (lower roughness), by cleaving the graphene sheet along its high symmetry crystallographic directions (armchair, zigzag). The as-defined graphene nanoconstrictions allow the observation of conductance quantization down to a few conductance quanta, even on standard $SiO_2$/Si substrates, at temperatures up to 40 K.

## Results and discussion

Graphene samples have been prepared by micro-mechanically exfoliation on the most commonly used silicon substrate with 285 nm $SiO_2$ capping layer (see Supplementary Materials S1). For defining high-quality graphene constrictions with high precision, we have developed an improved AFM-based lithographic technique (**fig. 1a**). To enable this, we have exploited the capability of AFM to determine the crystallographic orientations of the graphene lattice before patterning[32]. By imaging the surface in contact mode, the atomic potentials modulate the friction forces between tip and sample, revealing the high symmetry lattice directions of the graphene sheet (**fig. 1b, 1c**). This enables us to choose the precise crystallographic orientation of the cutting direction, which - as it turns out - is of key importance for defining high quality edges. For cutting the graphene sheet, the AFM tip is lowered and pushed into the sample surface to a predetermined deflection (corresponding to 100-200 nm apparent lowering of the tip, depending on the force constant), with the feedback-loop switched off. Then, the tip is moved along the line of the desired cutting direction (**fig. 1d**). Considering the unique tensile strength of graphene, for low indentation depths graphene is only stretched, adhering to the surface of the plastically deformed underlying silica substrate[33]. After reaching the indentation depth of approximately 1.5 nm, the chances of graphene rupture increase dramatically, in accordance with our previous findings[34].

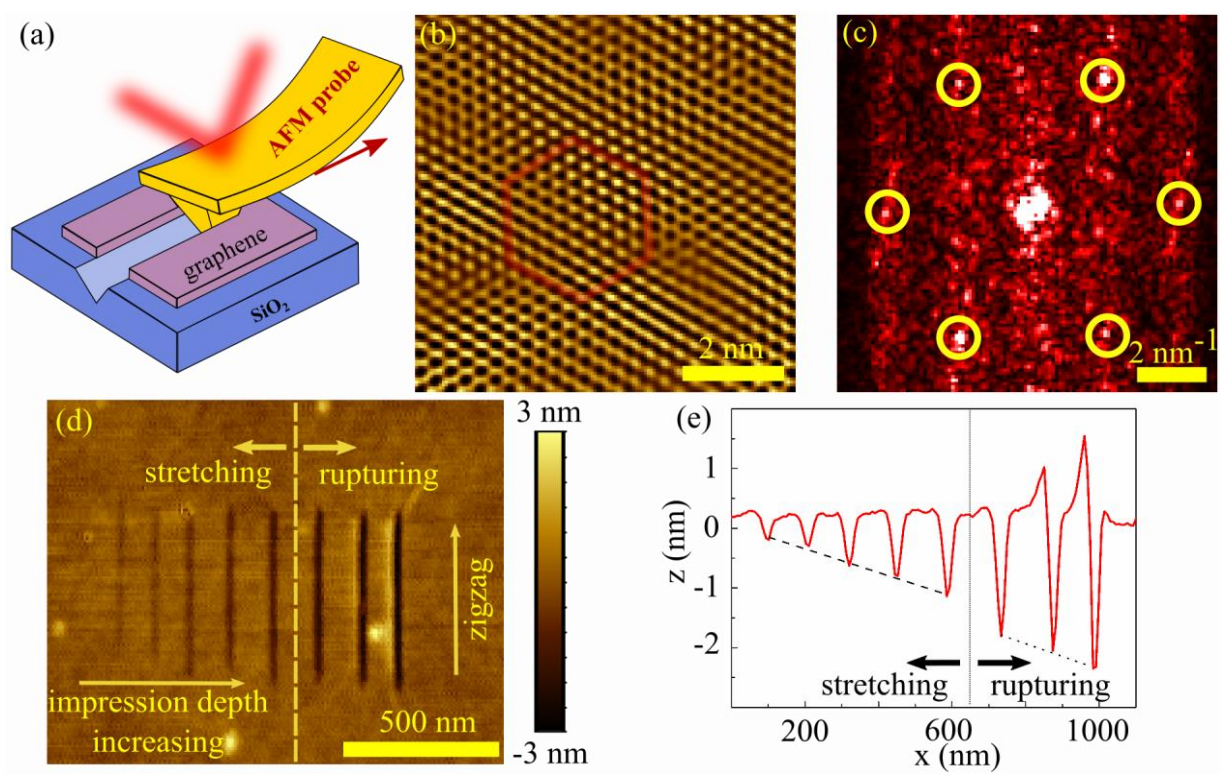

**Figure 1.** *AFM-based in-plane mechanical cleavage of graphene. (a) Schematic of the AFM cleavage lithography process. (b) Lattice resolution on graphene acquired in contact mode,*

*(FFT filtered for clarity), revealing the main crystallographic directions. (c) Fourier transform of the original lattice resolution image. (d) Linear AFM indentations in graphene along zigzag direction with gradually increasing depth. e) Depth profiles of the patterned lines from panel d, revealing the border between stretching and cutting of the graphene sheet.*

On fig. 1d, linear patterns running along a zigzag direction can be seen, prepared with gradually increased indentation depth. As a result of the smooth edges of the indentation line, it is hard to differentiate, when the graphene layer is cut instead of just being pushed/stretched into the groove of the substrate. However, by analyzing the line profiles (**fig. 1e**) the two regions can be distinguished. Further confirmation of the successful cutting of graphene comes from Raman spectroscopy mapping, where the D peak intensity is highly sensitive to the presence of edges (see Supplementary Materials S3).

Graphene constrictions can be formed by cutting two lines with one of their ends in close proximity, and the other ends reaching the edges of the flake. This, results in a narrow bridge between two large and intact graphene areas. To demonstrate the precision of our nanolithography technique in fig. 2 we show the topography of an approximately 10 nm wide constriction with the cutting lines running along an armchair direction. Note that while the two cutting lines have the same (armchair) orientation relative to the graphene lattice, they are not perfectly aligned. This slight misalignment helps avoiding strong backscattering, as discussed later. The smoothness of the edges is due to the fact, that graphene can be cut (cleaved) more easily along its high symmetry (armchair/zigzag) lattice directions[35,36] (see also Supplementary Materials S4). This is similar to cleaving bulk crystals along their high symmetry planes. Exploiting the unique ability of AFM to image and cut graphene along these easy cleaving directions, enables us to highly improve the precision of the nanofabrication process and reduce the roughness of edges. While the fabrication of graphene constrictions with ~ 10nm width is still challenging, constrictions of a few tens of nanometers can be reliably defined by our AFM cleavage lithography technique. To quantify the characteristic edge roughness, the phase image obtained in tapping mode proved to be the most suitable, as it reveals the boundaries more clearly than topographic images. The graphene edge – indicated by the black line in fig. 2c – is determined based on the deviation from the average value of the phase on the graphene sheet (44.4°). By analyzing the edge line, about +/- 1 nm deviations can be measured from the average $y$ position, indicating a ~ 2 nm edge roughness. This roughness is similar to that defined by cryo-etching of encapsulated graphene[23], but achieved using a far simpler technique. Most importantly, our AFM based nanofabrication technique has the advantage of avoiding energetic beams and aggressive chemical etching that can induce additional disorder extending tens of nanometers inwards from the nominal edges[19]. Moreover, the quality of the edges is further preserved, since after the mechanical cleavage, constriction edges never come into contact with resist material or wet chemistry.

Residues generated by material removal during the AFM cleavage lithography, stick to the AFM tip. At the end position of the cutting, the tip often deposit debris (see Supplementary Materials S2). To make sure that the leftover debris does not short circuit the constriction additional devices have been prepared, cutting across the whole graphene flake by AFM cleavage. The two-point resistance of such devices was in MΩ range, which is about two orders of magnitude higher than corresponding to conductance quanta, measured across constrictions, proving that the conductance of the cut section can be safely neglected.

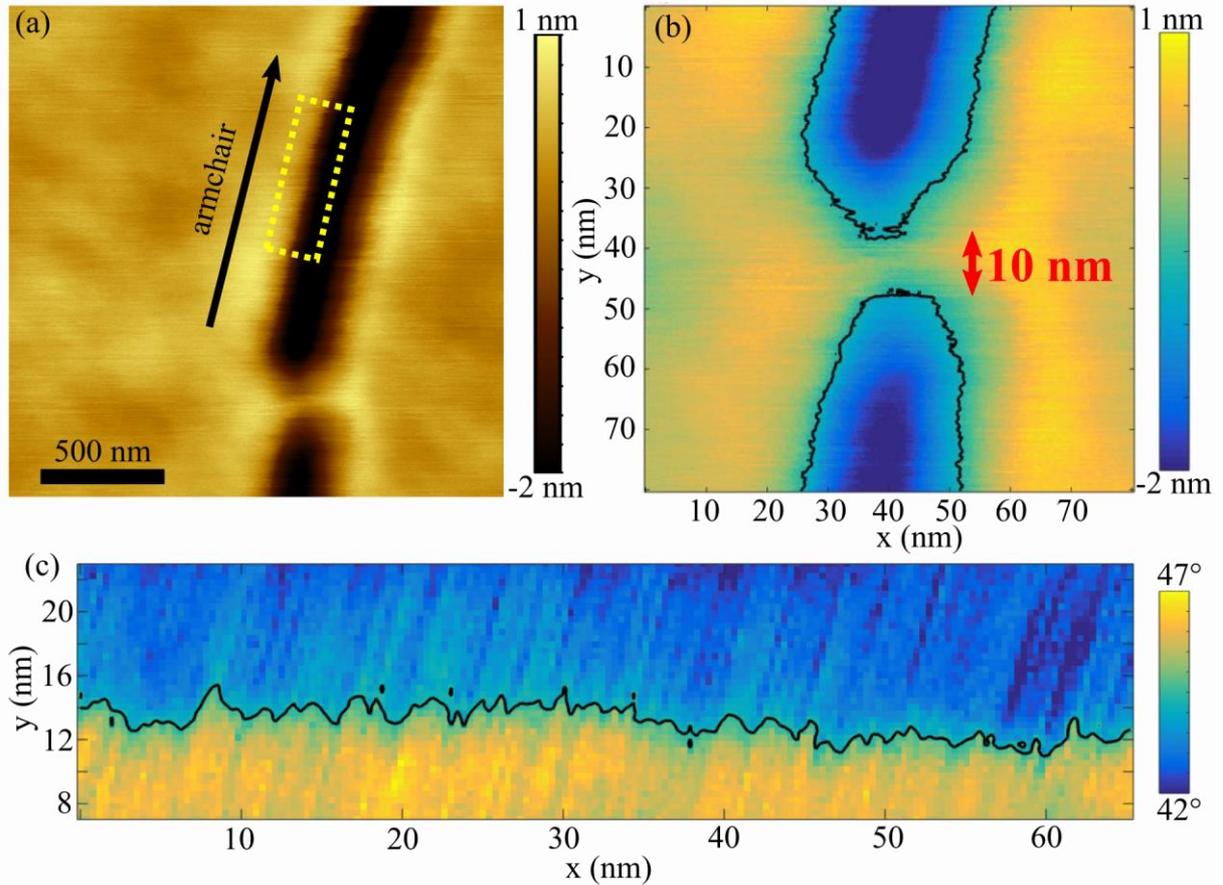

**Figure 2.** *High resolution AFM cleavage lithography of graphene quantum point contacts. (a) Topographic tapping-mode AFM image of a 10 nm wide graphene constriction. (b) Locating the graphene edges. (c) Estimating the edge roughness from the phase image of the dashed rectangular area of fig. 2a: The black line marks the border of the step-like deviation from the average phase value on graphene.*

Ti/Au (5/70 nm) electrodes have been defined on the graphene flakes by standard electron beam lithography technique prior to the AFM patterning step, to avoid the contamination of the constriction edges. To achieve lattice resolution AFM images, the resist residue has been eliminated by annealing in argon/hydrogen atmosphere at 400 °C [37]. In addition, an AFM sweeping-based cleaning technique was also applied (see Supplementary Materials S5.)[38,39]. Transport measurements in two-probe configuration have been performed at 1.5 K, using the heavily doped Si substrate as back-gate electrode. Plotting the conductance as a function of the applied gate voltage (**fig. 3a**) for a ~ 75 nm wide constriction patterned by AFM cleavage lithography, reveals well-defined plateaus around G values of 3, 5, 7, 9, and 11 e²/h, evidencing the conductance quantization, on top of the square root gate-dependence (see also Supplementary Materials S6).

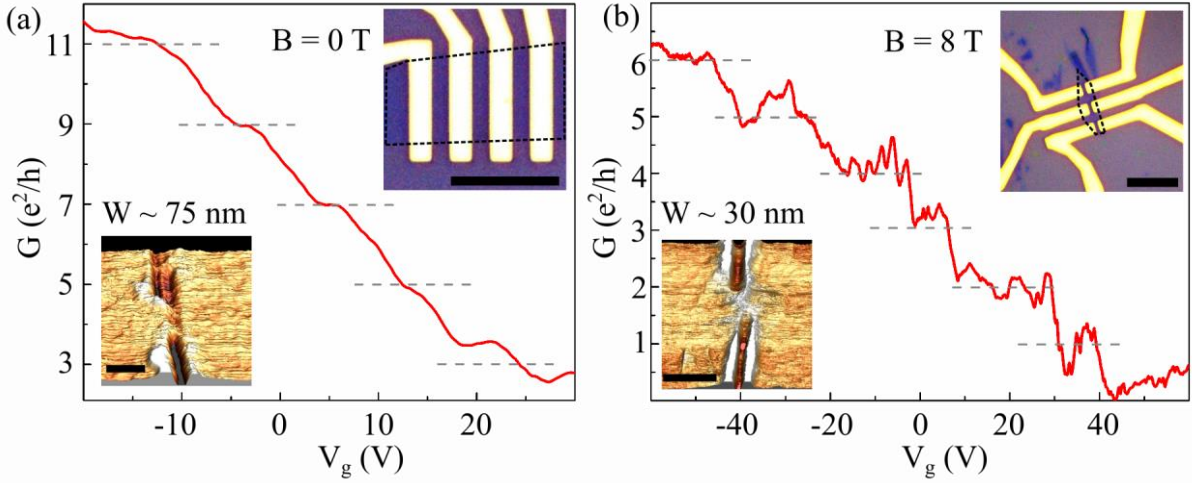

**Figure 3.** *Conductance measurements (1.5 K) of graphene QPCs patterned by AFM cleavage lithography with cuts running along zigzag directions. (Upper insets: optical images of the devices. Lower insets: 3D topographic AFM image of the nanoconstrictions, scale bars are 100 nm.) (a) Two-probe conductance curve in zero magnetic field, for a ~ 75 nm wide constriction revealing conductance plateaus at integer n∗e²/h values separated by 2∗e²/h steps. (b) Two-probe conductance curve measured through a ~30 nm wide graphene constriction in 8 T magnetic field displaying quantization steps with e²/h spacing.*

From the square root fitting of the G ($V_g$) characteristics, the width of the constriction can also be estimated. Using degeneracy factor of 2, and lever arm of $\alpha \approx 7.2*10^{10}$ cm$^{-2}$V$^{-2}$, a channel length of W ≈ 60 nm is obtained, which is in reasonably good agreement with the topographic AFM measurement (W ≈ 75 nm) of the constriction. Similarly good agreement have been found for the other investigated constrictions (see Supplementary Materials S6), supporting the interpretation of our results. For narrow constrictions, contact resistances can be safely neglected (see also Supplementary Materials S6). This is also confirmed by the fact that the conductance plateaus align reasonably well with $n \times e^2/h$ values even without contact resistance corrections (fig. 3). For graphene nanoribbons, conductance steps of 4 $e^2/h$ height are predicted[40], where the factor of 4 comes from the spin and valley degeneracy of the charge carriers. However, as also reported earlier, graphene constrictions often display 2 $e^2/h$ conductance steps instead[18,19,20,21], attributed to the lifting of the valley degeneracy. This is also predicted by the theoretical calculations of Guimarães *et al.* in which plateau-like features with a spacing of 2 $e^2/h$ can be observed in graphene nanoconstrictions that are less or equal in length than width ($L \leq W$)[41]. Additional constrictions fabricated by us by AFM lithography, display similar 2 $e^2/h$ steps, as can be seen in the Supplementary Materials section S8. We have also performed transport measurements with an externally applied (8 T) magnetic field. The results shown in fig. 3b. reveal conductance steps with roughly $e^2/h$ step heights. This can be attributed to the lifting of the spin degeneracy in addition to the lifted valley degeneracy. The phenomenon is similar to that observed on QPC devices approaching the quantum Hall regime[42]. On the inset of fig. 3. the layout of the devices as well as the 3D topographic AFM images of the corresponding nanoconstrictions are shown. We note that constriction edges appear less regular than in fig. 2 mainly due to less than optimal imaging conditions in order to preserve the structural integrity of the nanoconstrictions. The differences in the $V_g$ values (*x* axis of fig. 3) for different devices originate from the differences of the Dirac point position. A substantial *p* doping of the samples originates from the charged impurities in the $SiO_2$ substrate, as well as the annealing step employed for cleaning the samples from resist residues. Due to the relatively high *p* doping levels of graphene in our devices, we were not able to measure the electron branch of the characteristics. However, the significant *p* doping, together with the slight misalignment of constriction ends, helps us approaching the adiabatic limit, and avoiding strong back-scattering, due to sharp changes of the constriction widths on the scale of the Fermi wavelength. We also note that we could not obtain reliable conductance characteristics for the narrowest (~10 nm wide) constrictions fabricated here, as they were broken very fast during the measurements. Nevertheless, constrictions as narrow as ~30 nm could be measured (fig. 3b) displaying quantized conductance plateaus. Most constrictions also display oscillations superposed on the step-like conductance characteristics. This is commonly observed in QPC devices, and is often attributed to transmission resonances from reflections at the two ends of the constrictions, as well as quantum interference form impurity scattering near the constrictions[8].

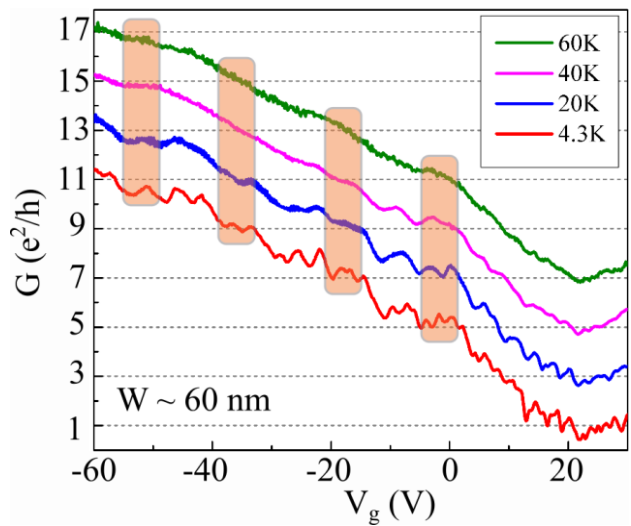

**Figure 4.** *Temperature dependence of QPC characteristics for a ~60 nm wide constriction cut along zigzag direction. For clarity the curves are successively offset by 2 $e^2$/h (from blue to green).*

We have also investigated the temperature dependence of the QPC characteristics (fig 4). In accordance with expectations, by increasing the temperature, the conductance plateaus are gradually smeared out, while also acquiring a finite slope, until they are no longer resolved. In our case, conductance plateaus could be clearly observed up to 20K, and signatures of conductance quantization could be detected even at 40K. This is in accordance with the estimated energy separation of transversal modes (~5 meV for 60 nm width), expected to persist up to ~$3k_BT$ (~5 meV at 20 K) thermal energy. Moreover, the energy separation of the transverse mode in the narrowest constriction fabricated by AFM lithography (order of 100 meV) is large enough for the QPC operation to persist even up to room temperature holding the promise of opening an entirely new regime for quantum point contact devices.

In summary, we have developed a high precision nanolithography technique for the fabrication of robust graphene quantum point contact devices. We showed that defining graphene constrictions through cleaving graphene sheets by AFM along their high symmetry directions, enables us to fabricate graphene nanoconstrictions down to 10 nm width and with reduced edge roughness of +/- 1 nm. Narrow graphene constrictions with high quality edges display robust QPC behavior manifesting in well-defined zero-field conductance quantization plateaus down to a few conductance quanta, even on $SiO_2$/Si substrates, and temperatures up to 40K. In zero magnetic field, evenly spaced conductance plateaus could be detected, roughly spaced by 2 $e^2$/h attributed to the lifting of the valley degeneracy, while the application of magnetic field resulted in plateaus spaced by $e^2$/h, indicating the lifting of both valley and spin degeneracy. Our novel AFM based nanopatterning technique enables the fabrication of robust graphene QPCs, lifting the highly demanding requirements for the device quality, prohibitive for most applications. Furthermore, such devices could also be employed as cheap resistance standards, providing precise resistance plateaus without requiring externally applied magnetic fields or very low temperatures for their reliable operation.

## Acknowledgements

This work has been supported by the NanoFab2D ERC Starting Grant, the Korea-Hungary Joint Laboratory for Nanosciences, and the Graphene Flagship, H2020 GrapheneCore3 project no. 881603. NIP acknowledges the support of the "Lendület" program of the Hungarian Academy of Sciences, LP2017-9/2017. The work was also supported by the NKFIH, Quantum Technology National Excellence Program 2017-1.2.1-NKP-2017-00001, by QuantERA

## Author contributions

L.T. and P.K conceived and designed the experiments. P.K. fabricated the samples and performed AFM and Raman measurements. I.E.L. supervised the e-beam lithography and evaporation processes. With the supervision of S.C., B.F., and P.K. carried out the electronic transport measurements. L.T. and C.H. supervised the project. P.K. analyzed the data and wrote the manuscript. L.T., G.D. and P.N.-I. advised on experiments, data analysis and writing the paper. All authors discussed the results and commented on the manuscript.

## Methods

Graphene sheets have been prepared by mechanical exfoliation of natural graphite crystals, with the standard scotch tape method on Si/SiO$_2$ substrate. Standard electron beam lithography (EBL) technique, followed by metal evaporation, has been utilized for device fabrication. To contact the graphene layer, Ti/Au (5/70 nm) has been evaporated at $10^{-8}$ mbar. To improve the quality of the devices, a 1-hour long annealing step has been performed in Ar:H$_2$ atmosphere at 400 °C. Additional to this, cleaning of graphene by AFM sweeping has been carried out (see Supplementary Materials S5). For AFM imaging and patterning, a Bruker Multimode 8 AFM has been used, equipped with a closed loop scanner. For patterning, diamond-like carbon coated silicon AFM probes (Tap300DLC, Budget Sensors) have been used, with a nominal force constant of 40 N/m and a tip radius of 15 nm. The AFM cutting has been controlled by the NanoMan lithography software of Bruker. For imaging the edges, AFM tips of 2 nm nominal tip radius (SSS-NCH type, NanoWorld) have been employed. The contact-mode AFM measurements have been carried out with contact tip (DNP-S10) characterized by nominal force constant of 0.06 N/m, and a tip radius of 10 nm. Transport measurements have been performed at 1.5 K in two-probe configuration using standard lock-in techniques at f=177 Hz and AC excitation voltage $V_{AC} = 10$ µV with a Stanford SR-830 lock-in amplifier and external low-noise I/V converter. For Raman mapping, a confocal Raman microscope system (Witec Alpha 300 RSA) has been used with Nd:YAG laser (532 nm, 2 mW).

## Data availability

The authors confirm that the data supporting the findings of this study are available within the article. Related additional data are available on reasonable request from the authors.

## Competing Interests

The authors declare that there are no competing interests.

## References

[1] Novoselov, K. S. *et al.* Two-dimensional gas of massless Dirac fermions in graphene. *Nature* **438**, 197–200 (2005).

[2] Zhang, Y., Tan, Y.-W., Stormer, H. L., & Kim, P. Experimental observation of the quantum Hall effect and Berry's phase in graphene. *Nature* **438**, 201–204 (2005).

[3] Morozov, S. V. *et al.* Giant Intrinsic Carrier Mobilities in Graphene and Its Bilayer. *Phys. Rev. Lett.* **100**, 016602 (2008).

[4] Miao, F. *et al.* Phase-Coherent Transport in Graphene Quantum Billiards. *Science* **317**, 1530-1533 *(*2007*)*.

[5] Stampfer, C. *et al.* Tunable Graphene Single Electron Transistor. *Nano Lett.* **8**, 2378–2383 (2008).

[6] Baringhaus, J. *et al.* Exceptional ballistic transport in epitaxial graphene nanoribbons. *Nature* **506**, 349–354 (2014).

[7] Bischoff, D. *et al.* Localized charge carriers in graphene nanodevices. *Appl. Phys. Rev.* **2**, 031301 (2015).

[8] Beenakker, C. W. J., & van Houten, H. Quantum Transport in Semiconductor Nanostructures. *Solid State Phys.* **44**, 1–228 (1991).

[9] van Wees, B. J. *et al.* Quantized Conductance of Point Contacts in a Two-Dimensional Electron Gas. *Phys. Rev. Lett.* **60**, 848–850 (1988).

[10] Wharam, D. A. *et al.* One-dimensional transport and the quantisation of the ballistic resistance. *J. Phys. C* **21**, L209 (1988).

[11] Zhang, Y. *et al.* Direct observation of a widely tunable bandgap in bilayer graphene. *Nature* **459**, 820–3 (2009).

[12] Overweg, H. *et al.* Electrostatically induced quantum point contacts in bilayer graphene. *Nano. Lett.* **18**, 553–559 (2017).

[13] Lee, H. *et al.* Edge-Limited Valley-Preserved Transport in Quasi-1D Constriction of Bilayer Graphene. *Nano Lett.* **18**, 5961 (2018).

[14] Banszerus, L. *et al.* Observation of the Spin-Orbit Gap in Bilayer Graphene by One-Dimensional Ballistic Transport. *Phys. Rev. Lett.* **124**, 177701 (2020).

[15] Alden, J. S. *et al.* Strain solitons and topological defects in bilayer graphene. *Proc. Natl Acad. Sci. USA* **110**, 11256–11260 (2013).

[16] Nakaharai, S., Williams, J. R., & Marcus, C. M. Gate-Defined Graphene Quantum Point Contact in the Quantum Hall Regime. *Phys. Rev. Lett.* **107**, 036602 (2011).

[17] Bischoff, D., Libisch, F., Burgdörfer, J., Ihn, T., & Ensslin, K. Characterizing wave functions in graphene nanodevices: Electronic transport through ultrashort graphene constrictions on a boron nitride substrate. *Phys. Rev. B* **90**, 115405 (2014).

[18] Tombros, N. *et al.* Quantized conductance of a suspended graphene nanoconstriction. *Nature Phys.* **7**, 697–700 (2011).

[19] Han, M. Y., Özyilmaz, B., Zhang, Y., & Kim, P. Energy Band-Gap Engineering of Graphene Nanoribbons. *Phys. Rev. Lett.* **98**, 206805 (2007).

[20] Molitor, F. *et al.* Energy and transport gaps in etched graphene nanoribbons. *Semicond. Sci. Tech.* **25**, 034002 (2010).

[21] Terrés, B. *et al.* Size quantization of Dirac fermions in graphene constrictions. *Nat. Commun.* **7**, 11528 (2016).

[22] Somanchi, S. *et al*. From Diffusive to Ballistic Transport in Etched Graphene Constrictions and Nanoribbons. *Ann. Phys. (Berlin)* **529**, 1700082 (2017).

[23] Clericò, V. *et al.* Quantum nanoconstrictions fabricated by cryo-etching in encapsulated graphene. *Sci Rep.* **9** 13572 (2019).

[24] Caridad, J. M. *et al*. Conductance quantization suppression in the quantum Hall regime. *Nat. Commun.* **9**, 659 (2018).

[25] Weng, L., Zhang, L., Chen, Y. P., & Rokhinson, L. P. Atomic force microscope local oxidation nanolithography of graphene. *Appl. Phys. Lett.* **93**, 093107 (2008).

[26] Masubuchi, S., Ono, M., Yoshida, K., Hirakawa, K., & Machida, T. Fabrication of graphene nanoribbon by local anodic oxidation lithography using atomic force microscope. *Appl. Phys. Lett.* **94**, 082107 (2009).

[27] Morgenstern, M., Freitag, N., Nent, A., Nemes-Incze, P., & Liebmann, M. Graphene Quantum Dots Probed by Scanning Tunneling Microscopy. *Ann. Phys.* **529**, 1700018 (2017).

[28] Vasic, B. *et al.* Atomic force microscopy based manipulation of graphene using dynamic plowing lithography. *Nanotech.* **24**, 0153303 (2013)

[29] He, Y. *et al.* Graphene and graphene oxide nanogap electrodes fabricated by atomic force microscopy nanolithography. *Appl. Phys. Lett.* **97**, 133301 (2010).

[30] Tapasztó, L., Dobrik, G., Lambin, P., & Biró, L. P. Tailoring the atomic structure of graphene nanoribbons by scanning tunnelling microscope lithography. *Nature Nanotech.* **3**, 397–401 (2008)

[31] Magda, G. Z. *et al.* Room-temperature magnetic order on zigzag edges of narrow graphene nanoribbons. *Nature* **514**, 608–611 (2014).

[32] Lee, C. *et al.* Frictional Characteristics of Atomically Thin Sheets. *Science* **328**, 76–80 (2010).

[33] Lee, C., Wei, X., Kysar, J. W. & Hone, J. Measurement of the Elastic Properties and Intrinsic Strength of Monolayer Graphene. *Science* **321**, 385–388 (2008).

[34] Nemes-Incze, P. *et al.* Preparing local strain patterns in graphene by atomic force microscope based indentation *Sci. Rep.* **7**, 3035 (2017).

[35] Kim, K. *et al.* Atomically perfect torn graphene edges and their reversible reconstruction. *Nat. Commun.* **4**, 2723 (2013).

[36] Kotakoski, J., & Meyer, J. Mechanical properties of polycrystalline graphene based on a realistic atomistic model. *Phys. Rev. B* **85**, 195447 (2012).

[37] Ishigami, M., Chen, J. H., Cullen, W. G., Fuhrer, M. S., & Williams, E. D. Atomic Structure of Graphene on $SiO_2$. *Nano Lett.* **7**, 1643–1648 (2007).

[38] Goossens, A. M. *et al.* Mechanical cleaning of graphene. *Appl. Phys. Lett.* **100**, 073110 (2012).

[39] Lindvall, N., Kalabukhov, A., & Yurgens, A. Cleaning graphene using atomic force microscope. *J. Appl. Phys.* **111**, 064904 (2012).

[40] Peres, N. M. R., Castro Neto, A. H. & Guinea, F. Conductance quantization in mesoscopic graphene. *Phys. Rev. B* **73**, 195411 (2006).

[41] Guimaraes, M. H. D., Shevtsov, O., Waintal, X., & van Wees, B. J. From quantum confinement to quantum Hall effect in graphene nanostructures. *Phys. Rev. B* **85**, 075424 (2012).

[42] Zimmermann, K. *et al.* Tunable transmission of quantum Hall edge channels with full degeneracy lifting in split-gated graphene devices. *Nat. Comm.* **8**, 14983 (2017)

# Robust quantum point contact operation of narrow graphene constrictions patterned by AFM cleavage lithography

# Supplementary Materials

## S1. AFM characterization of graphene thickness

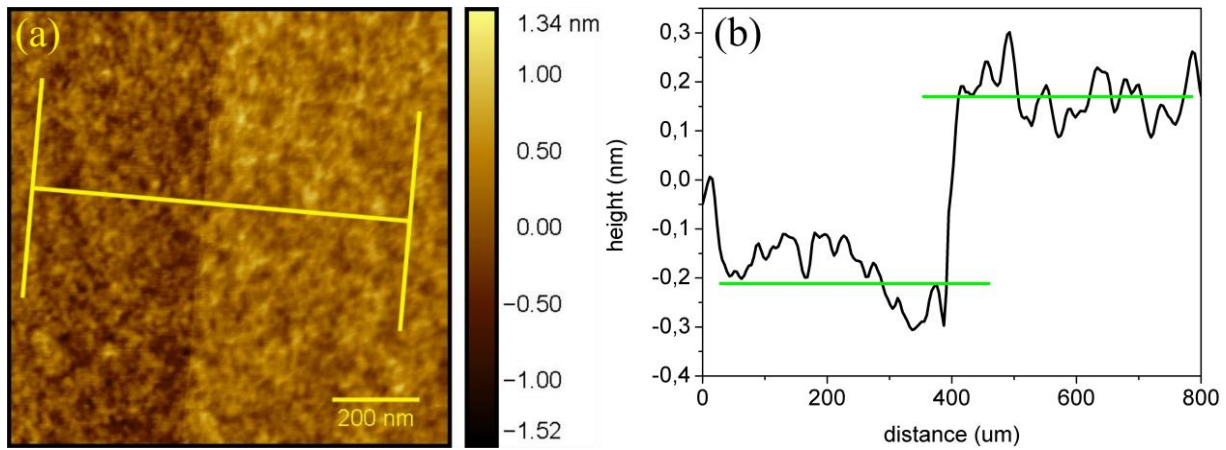

**Figure S1.** *Graphene thickness characterization. (a) Tapping mode AFM topography image of a natural edge of a graphene flake on Si/SiO$_2$ substrate. (b) averaged height profile of graphene (from the area shown in panel a) revealing a 0.4 +/-0.1 nm height, characteristic to single layers.*

AFM measurements confirm that the heights of the exfoliated graphene flakes are around 0.4+/- 0.15 nm relative to the SiO$_2$ substrate, evidencing their single-layer thickness. On figure S1a the AFM topography image of a natural edge of the graphene flake (right, brighter side) on Si/SiO$_2$ substrate (left, darker side) is shown. The height profile of the graphene edge (fig. S1b) is averaged over 128 scan lines, from the area marked in fig S1a, to reduce the noise, mostly due to the roughness of the underlying SiO$_2$ substrate, which is of the same order (root mean square roughness value of ~ 0.25 nm) as the thickness of the graphene.

## S2. Residues from the cleavage lithography

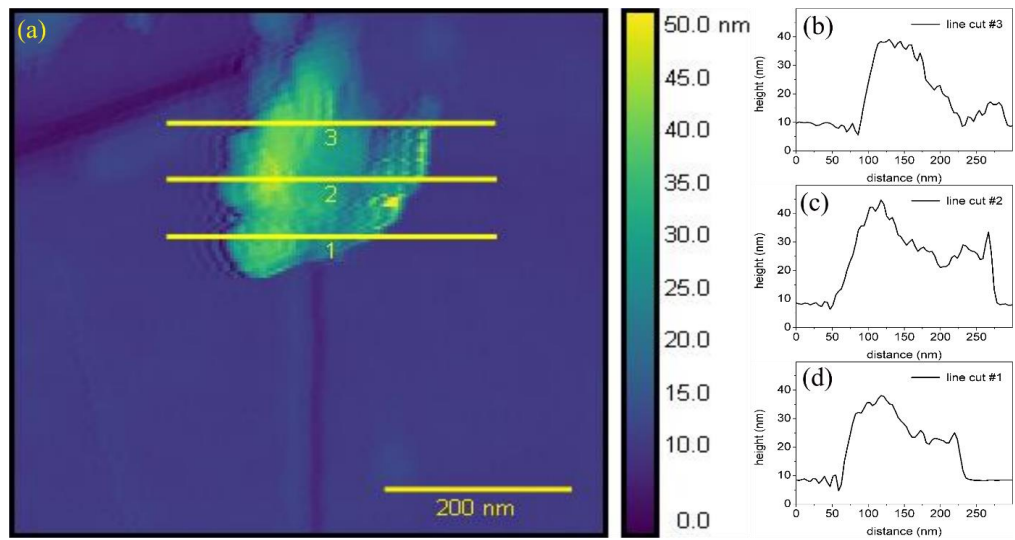

**Figure S2.** *Residues from the cleavage lithography. (a) Tapping mode AFM topography image of leftover debris located near the end of a cleavage lithography cut (b-d) height profiles of the debris along lines marked in panel (a)*

Our experience is that the material removed during the AFM cleavage lithography, mostly sticks to the AFM tip during the processes. At the end position of the cutting, the tip often deposits debris (fig. S2). That leftover residue is later swept away during scanning or it can be removed by AFM sweeping method (see also Supplementary Materials S5). The AFM tip can also be cleaned by performing indentations into the silica substrate. One particularly important question arises as to whether the graphene debris within the cutting lines could short circuit the constriction. To make sure this does not happen under the cutting conditions typically applied by us, we have performed test experiments, where we cut across the whole graphene flake by AFM cleavage. The two-point resistance of such test devices was always in MΩ range, which is about two orders of magnitude higher than corresponding to conductance quanta measured across constrictions. Therefore, the conductance of the cut section can be safely neglected.

# S3. Raman spectroscopy

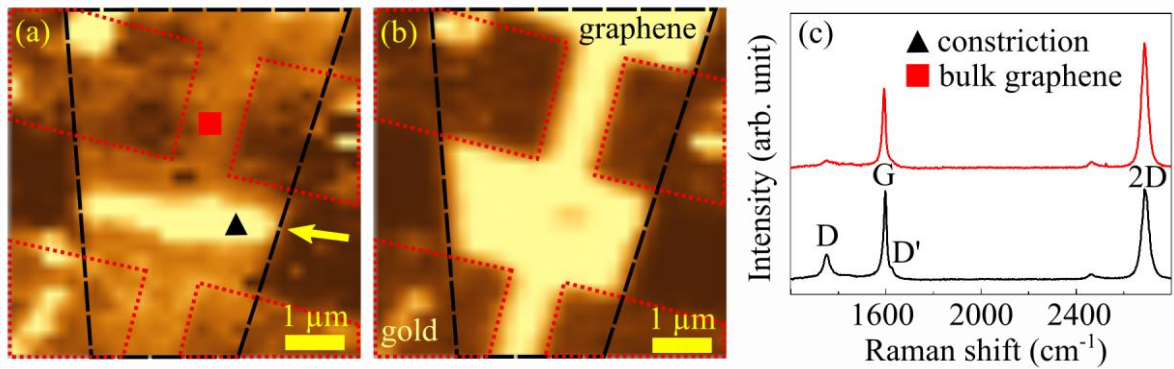

**Figure S3.** *Raman spectroscopy map of the D band (a) and G band (b) on contacted graphene comprising a constriction. (c) Raman spectra of graphene on the bulk area and around the constriction.*

Confocal Raman spectroscopy measurements have been carried out in order to investigate the results of the AFM lithography step. On fig. S3b the intensity map of the G band of graphene can be seen. The bright yellow area corresponds to the graphene flake with the presence of gold electrodes. Since the spot size of the excitation laser (~500 nm) is much larger than the width of the AFM defined trenches (~20 nm), only on the map of D band (fig. S3a) we can observe the signature of graphene edges defined by lithography[1,2]. Naturally, the small constriction near the middle of the line segment cannot be resolved in the Raman D map. Typical Raman spectra from the constriction area and from the pristine graphene area can be seen in fig. S3c. A clear sign for the presence of edges (successful cutting), is the $I_D/I_G$ ratio of 0.65, which is an order of magnitude higher around the constriction area than on pristine graphene (0.07). Aslo note the symmetric shape of the 2D peak that can be fit with a single Lorentzian, confirming the single-layer nature of the graphene flakes employed for QPC fabrication.

## S4. AFM lithography cutting along random lattice directions

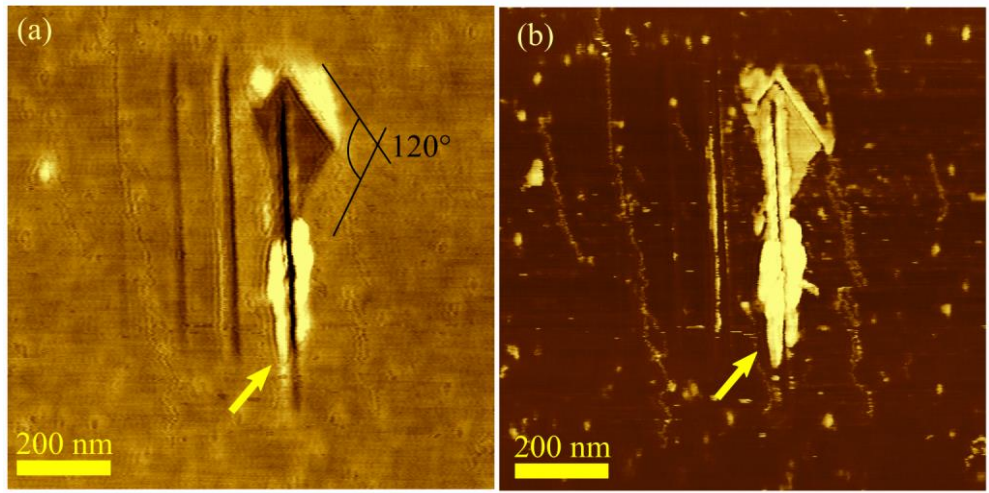

**Figure S4.** *Atomic force microscopy image of a series of indentations with random orientation in graphene. (a) Topography image in tapping mode; (b) phase image in tapping mode.*

The above indentation experiment is similar to the one presented in fig. 1d and discussed in the main text, but the lines were defined along random (not armchair, nor zigzag) orientations. Once graphene is cut (marked by green arrow), it develops rough edges (lower part). Then, at random points, it begins to tear along lines different from the cutting direction (AFM tip movement), in the upper part. The angles are multiples of 30 degrees in the corners, indicating that these easy tearing edges coincide with specific lattice directions (armchair, zigzag). This observation confirms the expectations that graphene tears more easily along special lattice direction (zigzag, armchair) that also implies smoother edge formation during cutting along these directions, similar to cleaving bulk crystals along their high symmetry planes.

## S5. Cleaning of resist residues by sweeping in contact mode AFM

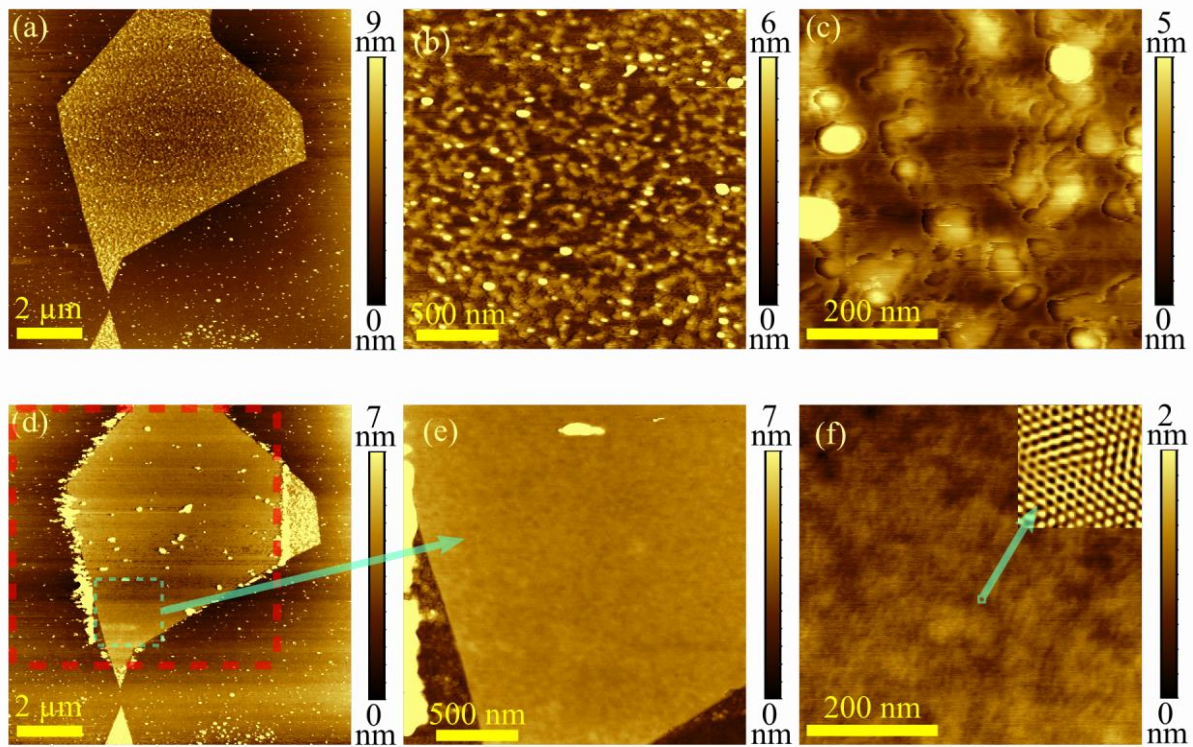

**Figure S5.** *Cleaning by AFM sweeping. (a-c) AFM topography images of a graphene flake covered with leftover resist material after the e-beam lithography step. (d-f) same flake after cleaning by AFM sweeping. Inset: of (f) shows the lattice resolution at the marked position.*

The mechanical cleaning based on contact mode AFM is a highly efficient cleaning technique at micro-and nanoscales[3,4]. The method relies on the careful establishment of the contact forces between the AFM tip and the surface, then continuous scanning (sweeping) of the sample surface with the appropriate set-point force. In fig. S5a-c atomic force topography images can be seen of a resist covered graphene flake. In this case the resist residues can be as high as 5 nm, leading to RMS roughness of 1.7 nm. Most of the contaminations are swept out after the first scan (fig. S5d, red rectangle). After multiple sweepings the graphene becomes clean, exhibiting an RMS values of only 0.2 nm. Showing the high efficiency of the cleaning. Even lattice resolution imaging becomes possible on such cleaned areas (fig. S5f).

## S6. Square root dependence of conductance

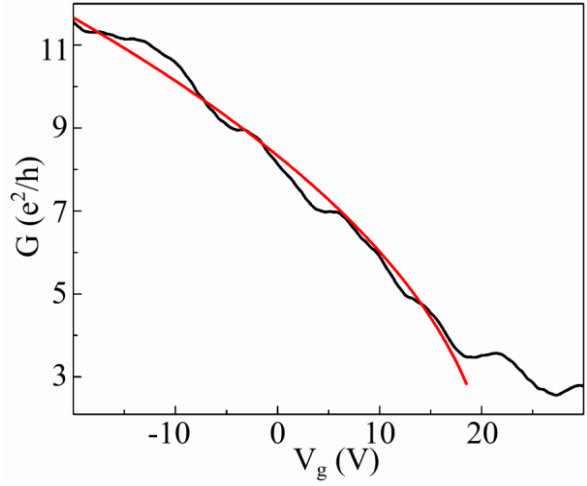

**Figure S6.** *Square root dependence of the conductance as a function of gate voltage through the graphene nanoconstriction of Fig 3a.*

The observation of the square root dependence is another proof that the electronic transport of the charge carriers can be described within the limits of the Landauer theory in our nanoconstriction devices. According to the Landauer theory the conductance through a W wide ideal nanoscale conductor increases by an additional $e^2/h$ conductance quantum as $Wk_F$ reaches a multiple of $\pi$ (an additional channel opens):

$$G = \frac{4e^2}{h} \sum_m \theta\left(\frac{Wk_F}{\pi} - m\right)$$

, where $\theta(...)$ is the Heavyside function, $k_F$ is the Fermi wave number and the factor 4 is due to the four-fold degeneracy of graphene. On the one hand, $k_F$ depends on the n carrier density: $k_F = \sqrt{\pi n}$. The applied back gate voltage ($V_g$) tunes the carrier density of the graphene layer as: $n = \alpha(V_g - V_D)$, where α is the so called lever arm, and $V_D$ is the charge neutrality point, the gate voltage at the minimal conductance. If we combine these equations:

$$G \propto \sqrt{n} \propto \sqrt{V_g - V_D}$$

On figure S4a a typical two-probe conductance measurement can be seen through graphene nanoconstriction. A good fit (R-square ~ 0.99) to square root dependence exists all the way down to G = 3 e$^2$/h.

The square root dependence of the conductance further confirms our assumption that the dominant contribution to the total resistance appears to be the resistance of the constriction at the scale of our examinations. The contact resistance of our devices was estimated to be below 500 Ω; therefore, it is negligible around the Dirac point where the resistance of the constriction is more than an order of magnitude higher.

From the square root fitting, the width of the constriction can also be estimated. Using degeneracy factor of 2 instead of 4 (as explained in the main text) and lever arm of α ≈ 7.2 × 10$^{10}$ cm$^{-2}$V$^{-2}$, a channel length of W ≈ 60 nm can be calculated which is in relatively good agreement with the topographic AFM measurement (≈ 75 nm) of the constriction. Further confirmation can be obtained by performing the square root fitting to the conductance measurements of additional devices. The measured channel length values (fig. S8a: 28 nm, fig. 4: 39 nm, fig. 3b: 17 nm) are in reasonably good agreement with the AFM measurements (fig. S8a: 45 nm, fig. 4: 60 nm, fig. 3b: 30 nm). Nevertheless, the extracted numbers indicate that topographic AFM measurements overestimate the constriction width by roughly 15-20 nm for all devices. This corresponds to about 8 nm error for each cut, which is about the practical resolution limit of typical AFM tips.

# S7. Charge carrier mobility estimation

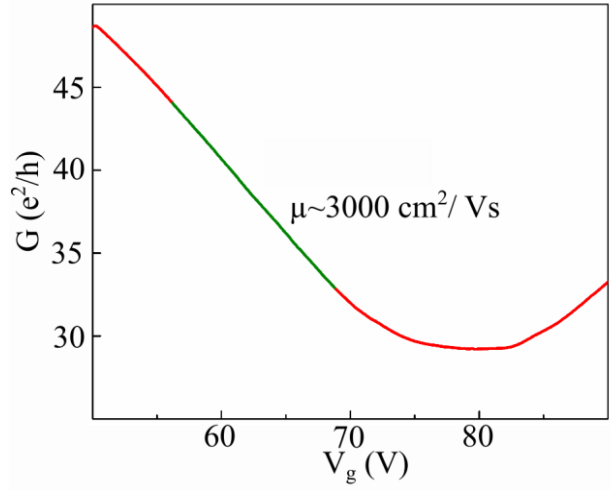

**Figure S7.** *Estimation of the charge carrier mobility of graphene from conductivity measurement on intact region of the same device as displayed in the main text (fig. 3a).*

From conductance measurements we can estimate the carrier mobility[5], as the back-gate voltage tunes the carrier density as:

$$n = \frac{\varepsilon_0 \varepsilon_r}{et} V_g$$

, where $\varepsilon_0$ and $\varepsilon_r$ are the permittivity of vacuum and $SiO_2$, e is the charge of the electron, and t is the thickness of the $SiO_2$ capping layer (285 nm). The carrier mobility ($\mu$) can be determined from the steepness of the conductance near the charge neutrality point (fig. S7), as:

$$\mu = \frac{G(V_g)}{en(V_g)}$$

The estimated mobility of most of our devices has been around 3000 $cm^2V^{-1}s^{-1}$, indicating a mean free path of 70 nm at a typical n ≈ $4*10^{12}$ $cm^{-2}$. As we measured constrictions of width from 30 to 75 nm, the scale of our constrictions is comparable to the estimated length of the mean free path.

## S8. Additional samples displaying conductance plateaus

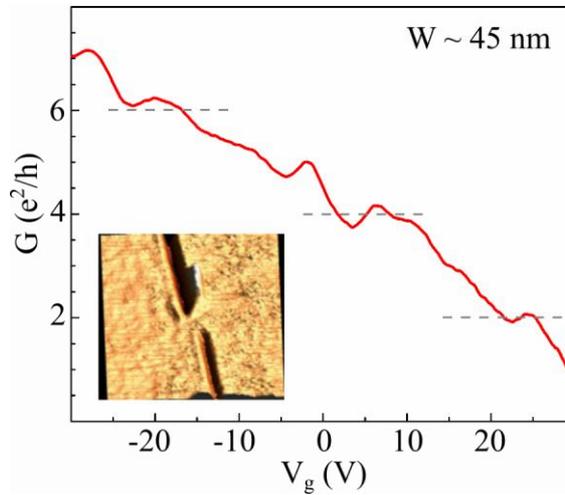

**Figure S8.** *Low temperature (1.5 K) two-probe conductance measurements at zero magnetic field revealing conductance quantization steps separated by 2 $e^2/h$, for a ~45 nm wide constriction cut along zigzag direction.*

Two-probe conductance measurements as a function of the back-gate voltage were performed on additional graphene nanoconstriction samples defined by AFM cleavage lithography. On figure S8 conductance plateaus can be detected at 2, 4, and 6 $e^2/h$. A step size of 2 $e^2/h$ can be observed on the QPC characteristics.

# Supplementary References

[1] Malard, L. M., Pimenta, M. A., Dresselhaus, G., & Dresselhaus, M. S. Raman spectroscopy in graphene. *Phys. Rep.* **473**, 51–87 (2009).

[2] Gupta, A. K., Russin, T. J., Gutiérrez, H. R. & Eklund, P. C. Probing Graphene Edges via Raman Scattering. *ACS Nano* **3**, 45–52 (2009).

[3] Goossens, A. M., et al. Mechanical cleaning of graphene. *Appl. Phys. Lett.* **100**, 073110 (2012).

[4] Lindvall, N., Kalabukhov, A. & Yurgens, A. Cleaning graphene using atomic force microscope. *J. Appl. Phys.* **111**, 064904 (2012).

[5] Novoselov, K. S. et al. Electric Field Effect in Atomically Thin Carbon Films. *Science* **306**, 666–669 (2004).